\documentclass[
reprint,
nofootinbib,
amsmath,amssymb,aps,
prl,
twocolumn,
]{revtex4-1}

\usepackage[utf8]{inputenc}
\usepackage{amsmath,graphicx,amssymb,bbold}
\usepackage{color}
\usepackage{mathtools}

\usepackage{graphicx}
\usepackage{dcolumn}

\begin{document}

\newcommand{\fd}[1]{{\color{blue}{#1}}}
\def\vec#1{\bm{#1}}
\def\ket#1{|#1\rangle}
\def\bra#1{\langle#1|}
\def\braket#1#2{\langle#1|#2\rangle}
\def\ketbra#1#2{|#1\rangle\langle#2|}

\preprint{APS/123-QED}

\title{Controlling Quantum Transport via Dissipation Engineering}

\author{Fran\c{c}ois Damanet$^{1}$, Eduardo Mascarenhas$^1$, David Pekker$^{2,3}$, and Andrew J.\ Daley$^{1}$}
\affiliation{$^{1}$ Department of Physics and SUPA, University of Strathclyde, Glasgow G4 0NG, United Kingdom.\\
$^{2}$ Department of Physics and Astronomy, University of Pittsburgh, Pittsburgh, Pennsylvania 15260, USA.
\\
$^{3}$ Pittsburgh Quantum Institute, Pittsburgh, Pennsylvania 15260, USA.
}

\date{\today}

\begin{abstract}

Inspired by the microscopic control over dissipative processes in quantum optics and cold atoms, we develop an open-system framework to study dissipative control of transport in strongly interacting fermionic systems, relevant for both solid state and cold atom experiments. We show how subgap currents exhibiting Multiple Andreev Reflections -- the stimulated transport of electrons in the presence of Cooper-pairs -- can be controlled via engineering of superconducting leads or superfluid atomic gases. 
Our approach incorporates dissipation within the
channel, which is naturally occurring and can be engineered in cold gas experiments. This opens opportunities for engineering many phenomena with transport in strongly interacting systems. As examples, we consider particle loss and dephasing, and note different behaviour for currents with different microscopic origin. We also show how to induce nonreciprocal electron and Cooper-pair currents.

\end{abstract}

\maketitle

\paragraph{Introduction.}

Understanding and controlling the out-of-equilibrium dynamics of strongly interacting many-body systems constitutes one of the key forefronts in quantum physics across a variety of subfields in experiment and theory. In this context, opportunities to achieve their control via dissipation mechanisms have arisen~\cite{Mul12, Dal14}, as is applied for few-body systems in quantum optics~\cite{Wis09, Gar15}. This is especially true in cold-atom platforms, where large separations between frequency scales allows well-controlled theoretical models and implementations of dissipative processes, as realized for laser cooling and trapping~\cite{Met01}. The longer timescales of cold atom experiments also allow dynamics to be tracked and potentially controlled time-dependently~\cite{Blo08, Blo12}. Out-of-equilibrium transport dynamics remain a ubiquitous paradigm in the solid state~\cite{Bee91}, and recent developments in cold atom systems have also made it possible to engineer quantised transport of atoms between reservoirs, as well as quantum point contacts and waveguides~\cite{Kri15,Kri17,Hus15,Leb18}. Here we explore the emerging new opportunity of using dissipation engineering to achieve control of quantum transport properties, that are relevant for both cold-atom and solid-state platforms.

We study transport in a system of strongly interacting fermions coupled to weakly interacting reservoirs, as can be reaslied with cold atoms using optical tweezers connecting larger superfluids, or with solid-state devices using quantum dots (QD) coupled to superconducting leads (S). In a traditional S-QD-S tunelling junction, subgap transport is known to be suppressed for weak electron tunneling as compared to the gap of the attached leads~\cite{Def10}. 
Here we demonstrate that subgap transport can be recovered even in the regime of weak tunnelling. This is done via reservoir engineering that allows for independent control of Cooper-pair and single-electron channels. Such channel separation can be accomplished in the solid state by adding two large-gap superconductors to a traditional S-QD-S junction, producing a four terminal structure, or in cold atoms considering driving from a molecular Bose-Einstein condensate~\cite{Jia11}.

Subgap currents in this context are produced by Multiple Andreev Reflections (MARs)~\cite{And64, Bee97, Mar11, Def10}, i.e., stimulated transport of electrons via exchange of Cooper-pairs. MARs have been observed in the solid state~\cite{Oct83, Pil10, Bui03, Dir11} and cold atoms~\cite{Hus15}, and their signatures can be used to reveal topological phase transitions related to Majorana bound states formation~\cite{San13}. We show how to engineer well-resolved MAR peaks under weak electron tunneling, and show how these behave in the presence of dissipation in the channel - providing a diagnostic tool for the microscopic nature of the current. We also show that for asymmetric coupling, the reciprocity of the engineered system is broken, yielding electron and Cooper pair currents dependent on the bias direction. This represents a genuinely new way of generating nonreciprocal transport of electrons and Cooper-pairs.

We investigate the transport properties of the junction with an open system approach, while most of the theoretical works rely on Keldish non-equilibrium Green functions or scattering techniques. These approaches are able to treat the tunnelling rate $\gamma$ between the QD and the leads non-perturbatively, but usually treat the Coulomb interaction $U$ between the QD electrons perturbatively or within a mean-field treatment~\cite{Yey97, Zaz06, Del08, Kan98}. In contrast, open system approaches such as input-output theories~\cite{Sea02, Gar04, Gar04b, Zha11} or master equations~\cite{Bre06, Kos13, Pfa13} work well in the opposite regime: for arbitrary interaction $U$ but weak tunnelling rate $\gamma$, implying that MARs have been left beyond their scope. In our framework, the large gap superconducting leads behave effectively as time dependent coherent drives of Cooper-Pairs on the QD (analogous to laser fields in quantum optics). This dynamical model is naturally cast as a dissipative Floquet system, for which we derive a Floquet-Born-Markov master equation~\cite{Yan16, Koh96, Gra94, Bla15} capturing MARs up to arbitrary order. Our open-system framework provides an opportunity to study the effects of controlled or uncontrolled dissipation acting on the QD. 
We thus analyze the response of the currents to fermion losses and dephasing, and show in particular robustness of the currents against dephasing.
We use in the following natural units in which $\hbar = k_B = e = 1$, where $-e$ is the electron charge.

\paragraph{Model.}

\begin{figure}[tb]
\begin{center}
\includegraphics[width=0.495\textwidth]{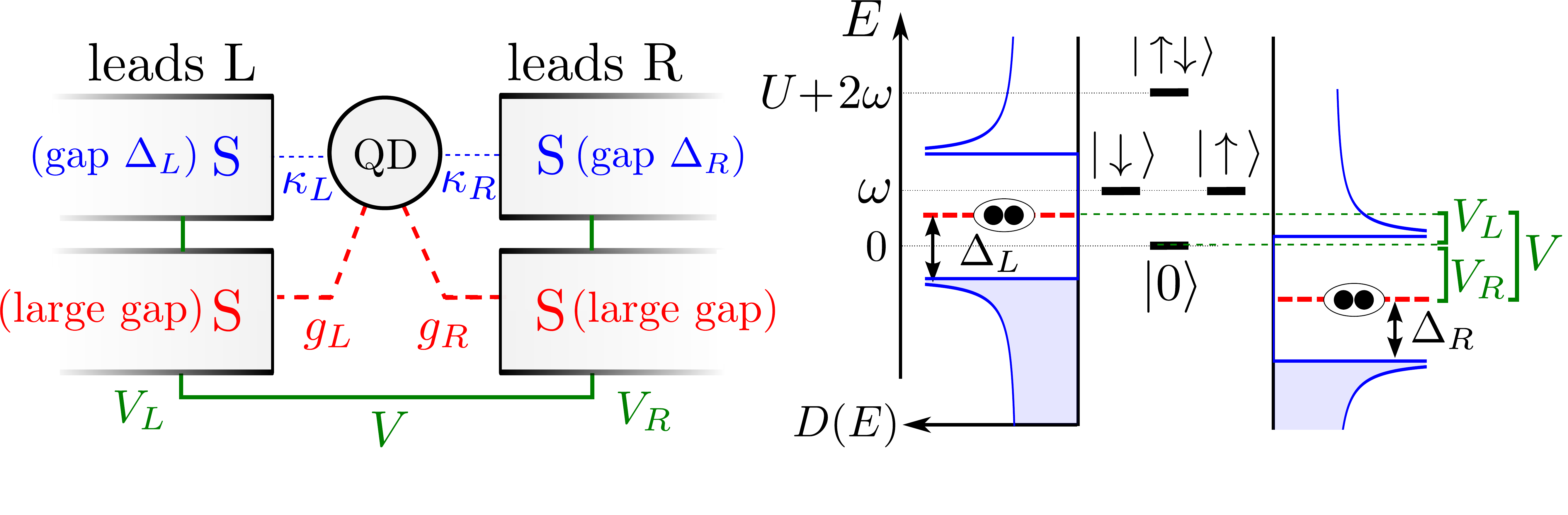}
\end{center}
\caption{(left) Sketch of the four-terminal QD tunnelling junction with Cooper-pair and electron tunnelling of amplitudes $g_\ell$ and $\kappa_\ell$ ($\ell = L,R$). (right) Corresponding energy diagram. The moderate-gap superconducting leads are characterized by Bogoliubov quasiparticles of density of states $D(E)$ (blue). The large-gap superconducting leads are only characterized by their Cooper-pair condensates (red). The applied bias voltage $V_\ell$ is the same for the moderate and large gap leads.}
\label{fig1}
\end{figure}

To represent the separate control of Cooper pair driving, we consider a four-terminal QD connected to two pairs of left (L) and right (R) superconducting leads by tunnel junctions, as depicted in Fig.~\ref{fig1}. In each pair, we consider one lead in the single-particle mean-field description with a moderate energy gap $\Delta_\ell$ ($\ell = L,R$), and one described only by its condensed fraction of Cooper-pairs, assuming that the gap is so large that single-particle excitations are irrelevant. A bias voltage $V = V_L - V_R$ is generated between the pairs of superconductors, where $V_L$ and $V_R$ are the voltages of each side. The QD Hamiltonian reads
\begin{equation}
H_\mathrm{QD} = \sum_{s = \downarrow, \uparrow} \omega c_s^\dagger c_s + U c_\uparrow^\dagger c_\uparrow c_\downarrow^\dagger c_\downarrow,
\end{equation}
and describes electrons of spin $s$, energy $\omega$, and Coulomb interaction $U$. The QD is an effective 4-level system spanned by the non-occupied, single occupied, and double-occupied states $\left\{\ket{0},\ket{\downarrow},\ket{\uparrow},\ket{\downarrow\uparrow}\right\}$. The coupling of the QD to the large-gap superconducting leads (red superconductors in Fig.~\ref{fig1}) gives rise to a pairing of the QD electrons, i.e., the proximity effect~\cite{Mar11}, and results in an effective time-dependent QD Hamiltonian of the form
\begin{equation}\label{HQDeff}
H_\mathrm{QD}^\mathrm{eff}(t) =  H_\mathrm{QD}  + \sum_{\ell = L,R}\big( g_\ell e^{2i V_\ell t}  c_\downarrow c_\uparrow + \mathrm{h.c.} \big),
\end{equation}
where $g_\ell$ is the Cooper-pair tunnelling amplitude between the QD and the large-gap superconducting $\ell = L,R$. Hence, the coupling of the large-gap superconductors with the QD takes the form of a driving of the transition between the non-occupied and double-occupied states $\ket{0}$ and $\ket{\downarrow \uparrow}$ of the QD.

We obtain the dissipative dynamics of the QD by coupling the Hamiltonian~(\ref{HQDeff}) to the superconductors with moderate gaps $\Delta_\ell$ (blue superconductors in Fig.~\ref{fig1}) under an open system approach, by deriving a Floquet-Born-Markov master equation~\cite{Yan16, Koh96, Gra94, Bla15} for the QD. The leads, considered in a mean-field single-particle description, act as baths of Bogoliubov quasiparticles of density of states $D_\ell(E) \propto \Theta(|E| - \Delta_\ell)|E|/(\sqrt{E^2 - \Delta_\ell^2})$. The tunnelling of electrons between the leads and the QD is described by a standard tunnelling Hamiltonian of the form $H_\mathrm{int} = \sum_{\ell}\kappa_\ell\sum_{ks} ( b^\dagger_{\ell ks} c_s + \mathrm{h.c.})$, where $\kappa_\ell$ is the electron tunnelling amplitude and $b_{\ell ks}$ the annihilation operator of an electron of spin $s$ and momentum $k$ in the moderate-gap lead $\ell$. The derivation of the master equation, in second-order in $H_\mathrm{int}$, results in a single-particle tunnelling rate $\gamma_\ell \propto \kappa_\ell^2$ (the typical line widths of the QD levels) considered as the smallest parameter. Note that while treating perturbatively the single-particle coupling, the master equation describes the QD Coulomb interaction $U$ exactly, as in~\cite{Kos13, Pfa13}. See Supplemental Material for details of the derivation~\cite{supmat}.

\paragraph{Engineering of transport.} 

\begin{figure}[tb]
\begin{center}
\includegraphics[width=0.48\textwidth]{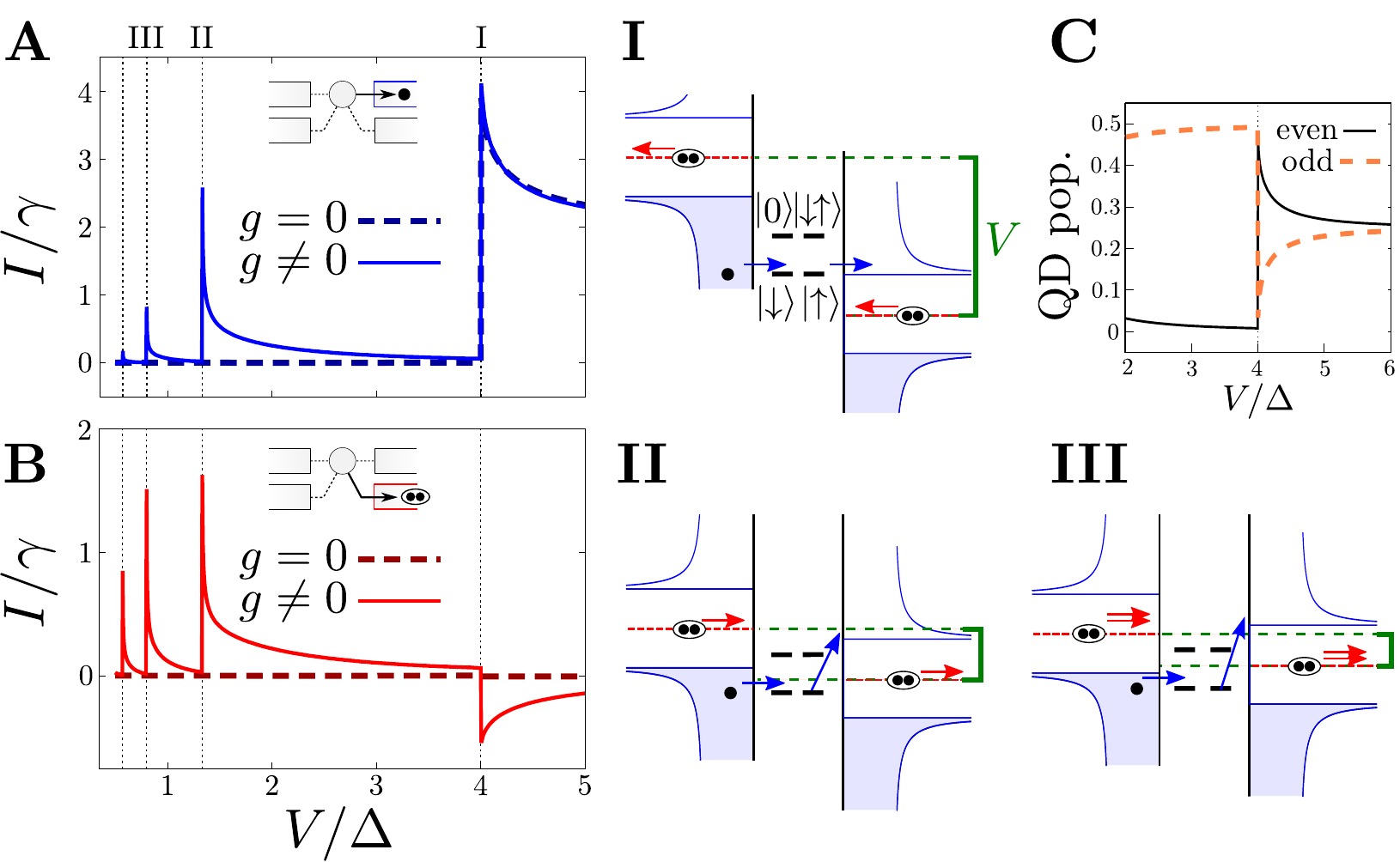}
\end{center}
\caption{Particle current $I$ (in units of $\gamma$) in the right moderate-gap lead (\textbf{A}) and large-gap lead (\textbf{B}) as a function of the bias voltage $V$ for $g = g_\ell = 0$ (dashed line) and $0.5$ (solid line). Other parameters are $U = 2$, $\omega = -1$, $\gamma = 10^{-2}$ and $T_\ell = 0$, in units chosen so that $\Delta_\ell \equiv \Delta = 1$. Subgap currents via MARs appear for non-vanishing $g$. (\textbf{I-III}) Energy diagrams corresponding to the standard resonant tunnelling (\textbf{I}) and the first and second MARs (\textbf{II} and \textbf{III}). \textbf{C}: Steady states QD averaged populations for states with even ($\ket{0},\ket{\downarrow \uparrow}$, black line) and odd ($\ket{\downarrow},\ket{\uparrow}$, dashed orange line) number of electrons.}
\label{fig2}
\end{figure}

From the solutions of the master equation, we calculate the particle current in the leads as a function of the applied bias voltages, taken as opposite from each other for the sake of simplicity ($V_L = -V_R = V/2$). Figure~\ref{fig2} shows the particle current $I$ in both the moderate and large-gap right leads as a function of $V$. We consider the electron-hole symmetric case $\omega = -U/2$, and vanishing or not Cooper-pair tunnelling $g \equiv g_\ell$, taken here real and identical for left and right leads. We also consider identical single-particle tunnelling rates $\gamma \equiv \gamma_\ell$ between the QD and the moderate-gap superconducting leads.

When $g = 0$, the large-gap superconducting leads are disconnected from the QD, and our system simply consists in a conventional S-QD-S tunnelling junction~\cite{Kos13,Pfa13}. Only one peak of current is observed (see panel A in Fig.~2), whose the shape is related to the superconductor density of states $D_\ell(E) \propto \Theta(|E| - \Delta_\ell)|E|/(\sqrt{E^2 - \Delta_\ell^2})$. The peak appears when $E_i > \omega > E_f$, where $E_i$ is the energy of the highest occupied state of the left lead and $E_f$ is the energy of the lowest non-occupied state of the right lead (see energy diagram I in Fig.~\ref{fig2}). For high bias, the particle current tends to the value $2\gamma$ of normal leads.
For low bias ($\gamma \ll V < 4\Delta_\ell$),  i.e.\ in the subgap region (where no resonance between left-lead occupied and right-lead non-occupied states exist), no current is observed as a result of the weak coupling approximation. Indeed, for $\gamma \ll \Delta$, Andreev reflection at the interface with the moderate-gap superconductor is negligible.

Connecting the large-gap superconducting leads to the QD (i.e., setting $g \neq 0$) allows Andreev reflections to occur. Under such process, an electron (hole) is reflected as a hole (electron) producing the emission (absorption) of a Cooper-pair in the large-gap superconducting leads (see panel B in Fig.~2). After some reflections, electrons of the QD acquire enough energy to tunnel into the moderate-gap superconducting lead. This produces well-resolved single-particle subgap currents more and more pronounced as $g$ increases.  These processes are represented in our Floquet-Born-Markov formalism by decay channels corresponding to QD transition (quasi)energies shifted by multiple of Cooper-pair energies (see Supplemental Material). The subgap currents are located at $V = 2(|\omega|+\Delta)/(2n+1)$ where $n = 1,2,\dots$ denotes the $n^{th}$ MAR (see energy diagrams II and III  corresponding respectively to the first and second Andreev reflections). This can be obtained from the condition $E_i + n V = \omega = E_f - n V$, in which $n$ denotes the number of Cooper-pairs transfer from the left to the right lead.
Note however that in general the bias voltage at which a MAR peak appears is a function of both the QD charging energy $U$ and $\omega$ (see Supplemental Material).
Hence, while the tunnelling between the QD and the moderate-gap leads is always sequential due to the weak-coupling regime (one electron per time), it can be assisted by transfer of an arbitrary number of Cooper-pairs between the large-gap superconductors, thanks to the stronger tunnelling amplitude $g$. This represents reservoir engineering of subgap single-particle currents. This is our first important result. Interestingly, the Cooper-pair current in the right large-gap superconducting lead is negative outside the subgap region. We attribute this phenomenon to a supercurrent (i.e., Cooper-pair current) reversal, due to the modification of parity of the QD when the voltage exceed the value delimiting the subgap border, as can be seen in the panel C in Fig.~\ref{fig2} \cite{Def10, Del18}. Note that the sign and amplitude of the supercurrent are dependent of the phases of the superconductors (not shown).

\paragraph{Effects of particle loss and dephasing.}

In the previous section, we showed that dissipation induced by reservoir engineering can be used to control subgap transport. Here we examine the robustness of the produced subgap currents against the presence of incoherent processes, that are inherent in real experimental setups. 
We incorporate these effects into our master equation through an additional dissipator of the form $\mathcal{D}_I(\rho) = \gamma_I \left(2 L \rho L^\dagger - \left\{ L^\dagger L, \rho \right\}\right)$, where $\gamma_I$ is the rate of the incoherent process and $L$ the corresponding Lindblad operator (see Supplemental Material). For cold atoms experiments, the dissipation in the channel is often in this Markovian form as can be derived from first principles~\cite{Dal14}.

\begin{figure}[tb]
\begin{center}
\includegraphics[width=0.475\textwidth]{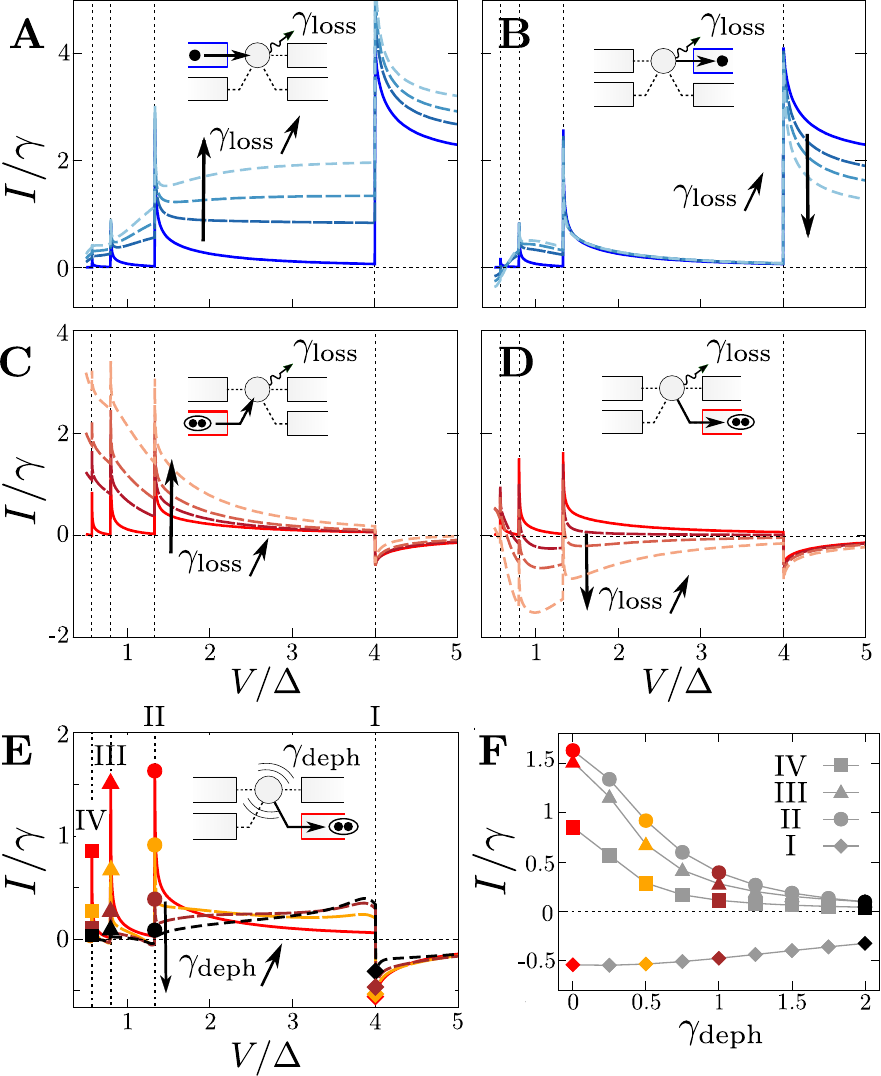}
\end{center}
\caption{Current-voltage characteristics under the effects of electron loss (\textbf{A-D}) and dephasing (\textbf{E-F}) acting on the QD. Currents of electrons entering (\textbf{A}) and leaving (\textbf{B}) the QD and of Cooper-pairs entering (\textbf{C}) and leaving (\textbf{D}) the QD as a function of the bias voltage $V$ for different electron loss rate $\gamma_\mathrm{loss}$ (from solid to dashed lines, $\gamma_\mathrm{loss} = 0$, $0.5 \gamma$, $\gamma$, $2\gamma$). 
Other parameters are $U = 2$, $\omega = -1$, $\gamma = 10^{-2}$, $T_\ell = 0$ and $g_\ell = 0.5$, in units of $\Delta_\ell \equiv \Delta$. Current of Cooper-pairs in the right large-gap superconducting lead as a function of $V$ for $\gamma_\mathrm{deph} = 0$, $0.5$,$1$ and $2$ (\textbf{E}); and as a function of $\gamma_\mathrm{deph}$ for $V = 2(|\omega|+\Delta)/(2n+1)$ with $n = 0$, $1$, $2$ and $3$ corresponding to the peaks I, II, III and IV (\textbf{F}). Other parameters as above.}
\label{fig3}
\end{figure}

We first consider the effects of particle loss (i.e., $\gamma_I \equiv \gamma_\mathrm{loss}$, $L = c_s$) acting on the QD. This occurs naturally in the cold-atom platforms through background gas collisions, and could be engineered using electron beams~\cite{Ger08} or light scattering quantum gas microscopes with single-site resolution~\cite{Wei11,Bak09,She10} (analogous to x-ray scattering in the solid state). In Fig.~\ref{fig3} (panels A-D), we show the particle currents in all the leads as a function of the bias voltage for increasing loss rates $\gamma_\mathrm{loss}$. The presence of losses results in competing effects. On the one hand, the additional decay channel tends to empty the QD faster. This results in an increase (decrease) of the currents of electrons entering (reaching) the moderate-gap superconducting leads. On the other hand, pushing the QD towards the non-occupied state $\ket{0}$ increases the effects of the driving (since the driving only affects the QD in the non-occupied or double-occupied states), which favors MARs and thus raises subgap currents. Hence, while source currents (panels A and C) only increase due to electrons losses, drain currents (panels B and D) are subjected to these competing effects, exhibiting amplitude increase or decrease depending on the voltage bias.

We then consider the effects of dephasing (i.e., $\gamma_I \equiv \gamma_\mathrm{deph}$, $L = c^\dagger_s c_s$) acting on the QD, which occurs naturally through coupling to additional degrees of freedom in the solid state, and can be engineered in cold atoms through light scattering or noise ~\cite{Dal14, Lus17,Sar14,Pic10,Nie17}. We show that dephasing acting on the QD affects identically the source and drain Cooper-pair currents, whereas leaves unchanged the electron currents. Figure~\ref{fig3} (panels E) shows the current of Cooper-pair leaving the QD to reach the right large-gap superconductor for different dephasing rate $\gamma_\mathrm{deph}$. Our results show that increasing the dephasing rate reduces the size of the subgap peaks (see panel F). This can be understood as a consequence of the blurring of the QD energy levels caused by the dephasing. Hence, dephasing tend to destroy Cooper-pair subgap currents, but does not affect the single particle currents. This suggests these latter are robust against phonon/photon scattering in condensed matter/cold atomic systems.

\paragraph{Nonreciprocal subgap transport.}

\begin{figure}[tb]
\begin{center}
\includegraphics[width=0.45\textwidth]{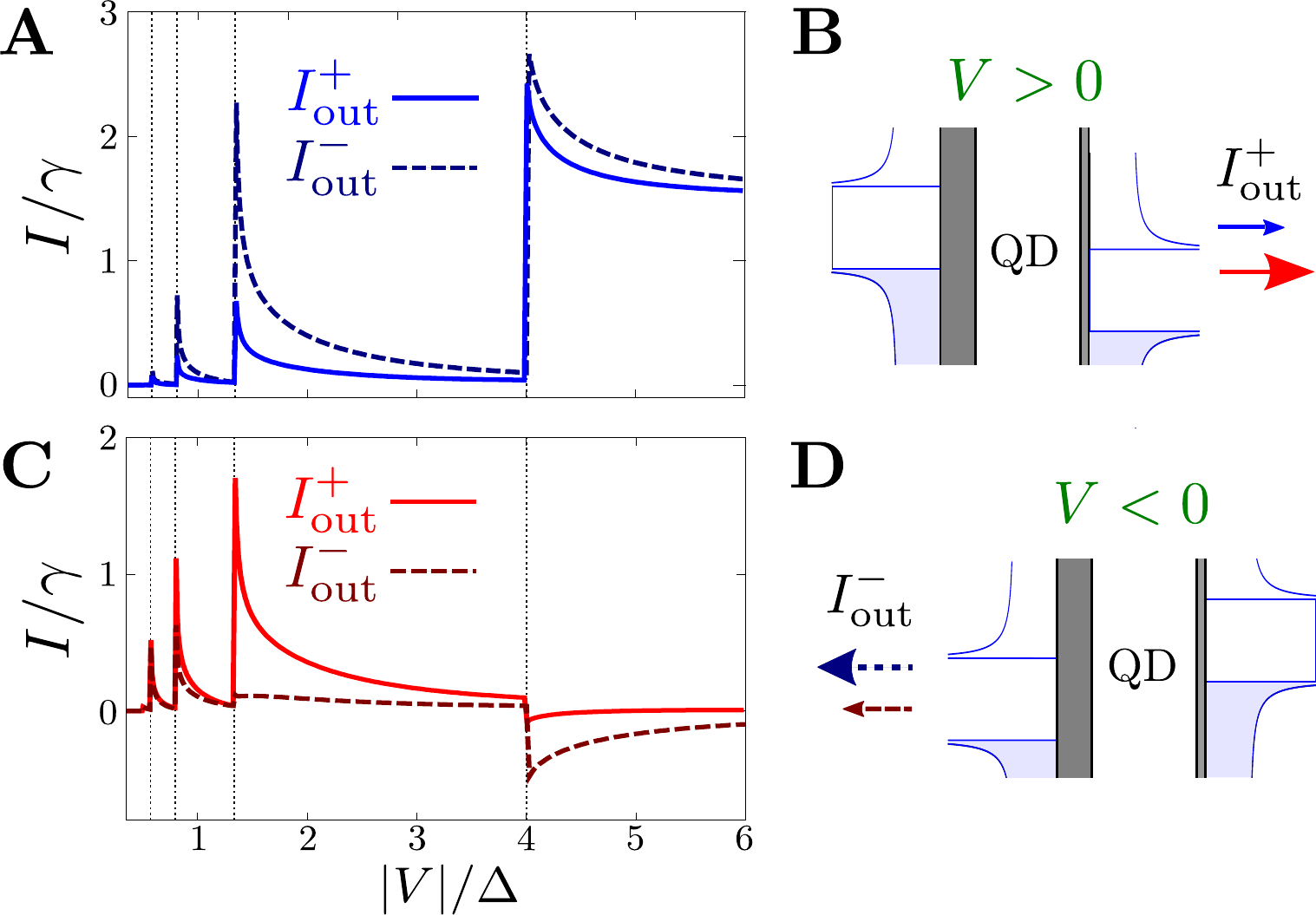}
\end{center}
\caption{Current-voltage characteristics in the moderate-gap (\textbf{A}) and large-gap (\textbf{C}) superconducting leads for asymmetric single-particle tunnelling rate $\gamma_L = 3 \gamma_R = 1.5 \; 10^{-2}$. In both plots, the solid (dashed) lines curves correspond to the current for positive (negative) bias voltage $V$, as depicted in the diagram \textbf{B} (\textbf{D}) on the right. Other parameters are $U = 2$, $\omega = -1$, $\gamma = 10^{-2}$, $T_\ell = 0$ and $g_\ell = 0.5$, in units of $\Delta_\ell \equiv \Delta$.}
\label{fig4}
\end{figure}

Finally, we show how to generate nonreciprocal subgap transport. Bias-direction-dependent properties is generally a desired feature of nanoscale devices, and are known to result from the presence of asymmetry and nonlinearity.  Nonreciprocal transport at the quantum level has been investigated in 
spin~\cite{Zala,Landi1,Landi,Emmanuel,Emmanuel2,Emmanuel3,Emmanuel4,Valente, Mas18} and QD systems~\cite{Mal18, Ono02, Joh05, Tan18, Sch08,Mal18}. This includes the paradigmatic Pauli Blockade effects in a double-QD junction, where a nonreciprocal electron current has been observed for asymmetric QD energy levels~\cite{Ono02}. 
For a single QD, the required asymmetry can be provided by different left and right tunnelling rates. In a S-QD-S junction, in the intermediate coupling regime ($\gamma_\ell \sim \Delta_\ell$), non reciprocal conductance has been observed and explained as originating from asymmetric Kondo resonance at the contact with the leads~\cite{Eic07}. 
Here we show that for asymmetric weak single-particle tunnelling rates $\gamma_L \neq \gamma_R$, the reciprocity of the transport properties can be broken as soon as the Cooper-pair tunnelling amplitudes $g_\ell$ is non-zero.
In Figure~\ref{fig4}, we plot the current-voltage characteristics for the moderate (panel A) and large gap (panel C) superconducting leads for positive and negative bias voltage (see diagrams B and D) for $\gamma_L = 3 \gamma_R$. While the total current (the sum of electron and Cooper-pair currents) is still reciprocal (not shown), its electron and Cooper-pair contributions become dependent of the bias direction, as can be clearly seen in Fig.~\ref{fig4}. In particular, the current of electrons (Cooper-pair) is larger (smaller) for negative (positive) bias. We interpret this phenomenon as a Cooper-pair-assisted nonreprocical transport, since it occurs only for non-zero $g_\ell$. Indeed, for $g_\ell = 0$, the electron current is reciprocal (not shown). We believe it is a genuinely new way of breaking reciprocity, since while keeping reciprocal the total current, its electron and Cooper-pair contributions -- which could be measured independently in our four-terminal scheme -- become asymmetric.

\paragraph{Conclusion.} 

We developed a quantum-optics-inspired framework to study the dynamics of strongly interacting fermions in tunnelling junctions under the influence of dissipation and driving, relevant for both solid-state and cold-atom platforms. For concreteness, we studied the dynamics of a QD coupled to superconducting leads in a four-terminal configuration, where two large-gap superconducting leads are added to a traditional S-QD-S tunnelling junction. We demonstrate the possibility of controlling subgap transport via dissipation engineering.
We showed that the added leads generate subgap transport based on MARs despite weak electron tunnelling, and studied the effects of electron loss and dephasing acting on the QD. Finally, we showed that the Cooper-pair driving provided by the added leads is a new way of breaking the reciprocity of the junction, generating nonreciprocal electron and Cooper-pair subgap currents based on MARs.

Our results could be investigated in both solid-state and cold-atom experiments. They could be generalized to multi-QD tunnelling junction, and to include the presence of measurement and feedback loop to control the transport dynamics of fermions in tunnelling junctions~\cite{Uch18}.
More possible outlooks include reservoir engineering of (Floquet)-Majorana fermions~\cite{Jia11, Ben15, Li014}, or studies of the interplay between dissipation and driving in thermodynamics problems such as thermoelectric effects~\cite{RSan13} or quantum heat engines~\cite{Pek15} involving superconductors.

\begin{acknowledgments} 
Acknowledgments: Work at the University of Strathclyde was supported by the EPSRC Programme Grant DesOEQ (EP/P009565/1), and by the EOARD via AFOSR grant number FA9550-18-1-0064. Work at the university of Pittsburgh was supported by NSF PIRE-1743717. 
\end{acknowledgments}

\begin{widetext}

\section{SUPPLEMENTAL MATERIAL}

In this Supplemental Material, we derive the Floquet-Born-Markov master equation used to investigate the non-equilibrium dynamics of our four-tunnelling junction. We then explain how to solve the master equation and compute the different particle currents, and how to incorporate the effects of additional incoherent processes acting on the QD. We provide some details of our numerical simulations, and finally show the dependence of the positions of the MAR peaks on the QD energy.

\section{Derivation of the master equation}

We model the junction with the Hamiltonian $H = H_{\mathrm{QD}}^\mathrm{eff}(t) + H_{\mathrm{lead}}+ H_{\mathrm{I}}$ where
\begin{equation}\label{model2}
\begin{aligned}
&H_\mathrm{QD}^\mathrm{eff}(t) =  \sum_{s = \downarrow, \uparrow} \omega c_s^\dagger c_s + U c_\uparrow^\dagger c_\uparrow c_\downarrow^\dagger c_\downarrow + \sum_{\ell}\big( g_\ell e^{2i V_\ell t}  c_\downarrow c_\uparrow + \mathrm{h.c.} \big),\\
&H_{\mathrm{lead}} = \sum_{\ell = L,R} \sum_{k} \Big[\sum_{s = \downarrow, \uparrow}\left(\omega_{k} + V_\ell \right) b_{\ell ks}^\dagger b_{\ell ks} + \sum_{k'} U_{kk'} b_{\ell k \uparrow}^\dagger b_{\ell k\downarrow}^\dagger b_{\ell -k'\uparrow} b_{\ell -k'\downarrow}  \Big], \\
&H_{\mathrm{I}} = \sum_{\ell = L,R}\kappa_\ell\sum_{ks} \left( b^\dagger_{\ell ks} c_s + \mathrm{h.c.}\right). \\
\end{aligned}
\end{equation}
In Eq.~(\ref{model2}), $H_\mathrm{QD}^\mathrm{eff}(t)$ is the effective time-dependent QD Hamiltonian which describes the intrinsic dynamics of the QD and the contribution of the left and right large-gap superconducting leads. The time-dependent pairing terms can be viewed as coming from an interaction Hamiltonian of the form
\begin{equation}\label{CPtunnelling}
\begin{aligned}
&H_{\mathrm{I}}^{(p)} = \sum_{\ell}\big( g_\ell S^\dagger_\ell c_\downarrow c_\uparrow + \mathrm{h.c.} \big),
\end{aligned}
\end{equation}
where $S_{\ell}$ is the annihilation operator of a Cooper-pair of energy $2V_\ell$ in the large-gap superconducting $\ell$, on which one applies a semi-classical description by considering the fields $S_\ell$ in coherent states, making the replacement $S_\ell \to e^{-2i V_\ell t}$, and rescaling $g_\ell$. Note that we consider here $g_\ell$ arbitrary, since it can be tuned by adjusting the properties of the large-gap superconducting leads and their coupling with the QD. The Hamiltonian $H_{\mathrm{lead}}$ describes the moderate-gap superconducting leads where $b_{\ell ks}$ is the annihilation operator of an electron of spin $s$ and energy $\omega_k$, $V_\ell$ the bias voltage, and $U_{kk'}$ the interaction between electrons. Finally, the Hamiltonian $H_{\mathrm{I}}$ describes electron tunnelling between the QD and the leads with amplitudes $\kappa_\ell$. Our four-terminal model allows us by essence to consider independent electron and Cooper-pair tunnelling. We note however that independent tunnelling can also be the result of modelling superconducting leads with two components, as done in~\cite{Su17}.

We consider the system-bath (S-B) decomposition $H_\mathrm{S}(t) \equiv H_\mathrm{QD}^\mathrm{eff}(t)$ and $H_\mathrm{B} \equiv H_\mathrm{lead}$, assuming the electron tunnelling amplitude $\kappa_\ell$ to be the smallest frequency scale. In order to take into account the time-dependence of the system Hamiltonian in the dissipative processes, we derive a Floquet-Born-Markov master equation, i.e., a Born-Markov master equation based on the Floquet quasienergy spectrum of $H_\mathrm{S}(t)$~\cite{Yan16, Koh96, Gra94}. This procedure (i.e. adding the time-dependent driving first and then calculating dissipation) is justified since the coupling with the large-gap superconductors is larger than the ones with the moderate-gap leads. This allows the dissipation to be assisted by Cooper-pair exchanges, yielding the subgap peak structures observed in the main text. The reverse procedure (i.e. calculating dissipation first and then adding the time-dependent driving) is unable to describe the interplay between the electrons and the Cooper-pairs, and is anyway not justified in our regime of parameters.

\subsection{Interaction picture and Bogoliubov approximation}

Our starting point is the usual Liouville-Von Neumann equation~\cite{Bre06}
\begin{equation}\label{LVN}
\dot{\rho}_\mathrm{tot}^I(t) = - i\left[ H_\mathrm{I}^I(t),\rho_\mathrm{tot}^I(t)\right] 
\end{equation}
for the total density matrix $\rho_\mathrm{tot}^I(t)$ in interaction picture with respect to $H_0(t) = H_\mathrm{S}(t) + H_\mathrm{B}$, where
\begin{equation}
 H_I^I(t) = \sum_\ell \kappa_\ell \sum_{k s} \left(b_{\ell ks}^{I\dagger}(t) c_s^I(t) +  \mathrm{h.c.} \right),
\end{equation}
with
\begin{align}\label{ctbt}
&c_s^I(t) =  U(t)^\dagger c_s U(t),\\
&b_{\ell ks}^I(t) = e^{i H_\mathrm{B} t} b_{\ell ks} e^{-i H_\mathrm{B} t},
\end{align} 
where the propagator $U(t)$ is defined as
\begin{equation}\label{propa}
U(t)= \mathcal{T} e^{- i \int_0^{t} H_\mathrm{QD}^\mathrm{eff}(t') dt'},
\end{equation}
with $\mathcal{T}$  the time-ordering operator. 

In order to calculate $b_{\ell ks}^I(t)$, we first evaluate the time-dependence due to the bias voltages $\sum_\ell V_\ell N_\ell = \sum_{\ell ks} V_\ell  b_{\ell ks}^\dagger b_{\ell ks}$ where $N_\ell$ is the total electron number operator in the lead $\ell$, which yields a factor $e^{-i V_\ell t}$. This can be done since $\sum_{\ell} V_\ell  N_\ell$ and $H_B - \sum_{\ell} V_\ell  N_\ell$ commute. We then obtain the time-dependence provided by the remaining Hamiltonian $H_B - \sum_{\ell} V_\ell  N_\ell$ by performing the Bogoliubov mean-field approximation on $H_B - \sum_{\ell} V_\ell  N_\ell$. For this purpose, we define the order parameter $\Delta_\ell e^{i \phi_\ell}= - \sum_k' U_{kk'} \langle b_{\ell -k'\downarrow} b_{\ell k' \uparrow} \rangle$ where $\Delta_\ell$ is the energy gap of the lead $\ell$ and then diagonalize $H_B - \sum_{\ell} V_\ell  N_\ell$ which yields
\begin{equation}
H_{\mathrm{B}} = \sum_{\ell} \sum_{ks} \left( \tilde{\omega}_{\ell k}\tilde{b}_{\ell ks}^\dagger \tilde{b}_{\ell ks} + V_\ell b_{\ell ks}^\dagger b_{\ell ks} \right),
\end{equation}
where $\tilde{b}_{\ell, ks}$ describes the annihilation of a Bogoliubov quasiparticle of spin $s$ and energy $\tilde{\omega}_{\ell k} = \sqrt{\omega_{k}^2 + \Delta_{\ell}^2}$ obtained from the Bogoliubov transformation
\begin{equation}
\begin{aligned}
&b_{\ell -k\downarrow} = u_{\ell k} \tilde{b}_{\ell k\downarrow} + v_{\ell k} \tilde{b}_{\ell k\uparrow}^\dagger, \\
&b_{\ell k\uparrow} = u_{\ell k} \tilde{b}_{\ell k\uparrow} - v_{\ell k} \tilde{b}_{\ell k\downarrow}^\dagger, \\
\end{aligned}
\end{equation}
where 
\begin{equation}
\begin{aligned}
&u_{\ell k} = e^{-i \phi_\ell} \sqrt{\frac{1}{2}\left(1+\frac{\omega_k}{\tilde{\omega}_{\ell k}}\right)}, \\
&v_{\ell k} = \sqrt{\frac{1}{2}\left(1 -\frac{\omega_k}{\tilde{\omega}_{\ell k}}\right)}.
\end{aligned}
\end{equation}
The time-dependence of the lead electron operators thus read
\begin{equation}\label{intpic}
\begin{aligned}
&b_{\ell k\downarrow}^I(t) =\left(u_{\ell k} \tilde{b}_{\ell -k \downarrow} e^{- i \tilde{\omega}_{\ell k} t} + v_{\ell k} \tilde{b}_{\ell -k \uparrow}^\dagger e^{i \tilde{\omega}_{\ell k} t}\right) e^{-i V_\ell t}, \\[8pt]
&b_{\ell k\uparrow}^I(t)= \left(u_{\ell k} \tilde{b}_{\ell k \uparrow} e^{- i \tilde{\omega}_{\ell k} t} - v_{\ell k} \tilde{b}_{\ell k \downarrow}^\dagger e^{i \tilde{\omega}_{\ell k} t}\right) e^{-i V_\ell t}.
\end{aligned}
\end{equation}

\subsection{Born and Markov approximation} 

We consider the Born approximation, i.e., the total system-bath density matrix $\rho_\mathrm{tot}^I(t)$ in the separable form 
\begin{equation}\label{Bornapp}
\rho_\mathrm{tot}^I(t) \approx \rho^I(t) \otimes \rho_L \otimes \rho_R
\end{equation}
where $\rho^I$ is the QD density matrix in interaction picture and $\rho_\ell$ a stationary state of the lead $\ell$, taken here as the thermal state
\begin{equation}
\rho_\ell  = \frac{e^{- \beta_\ell \left(H_B - V_\ell N_\ell\right)}}{\mathrm{Tr}\left[e^{- \beta_\ell \left(H_B - V_\ell N_\ell\right)}\right]} 
\end{equation}
so that the Fermi levels lie in the middle of the gaps of their respective leads, where $\beta_\ell = 1/k_B T_\ell$ with $T_\ell$ their temperature. In doing so, we obtain the simple expressions
\begin{equation}\label{thermal}
\begin{aligned}
&\langle \tilde{b}_{\ell ks}^\dagger \tilde{b}_{\ell  k^\prime s^\prime} \rangle_B \equiv \mathrm{Tr}_B\left[ \tilde{b}_{\ell  ks}^\dagger \tilde{b}_{\ell  k^\prime s^\prime} \rho_\ell \right] = \delta_{kk'}\delta_{ss'} n_\ell(\tilde{\omega}_{\ell k}), \\
&\langle \tilde{b}_{\ell  ks} \tilde{b}_{\ell k^\prime s^\prime}^\dagger \rangle_B \equiv \mathrm{Tr}_B\left[ \tilde{b}_{\ell ks} \tilde{b}_{\ell k^\prime s^\prime}^\dagger \rho_\ell \right] = \delta_{kk'}\delta_{ss'} (1 - n_\ell(\tilde{\omega}_{\ell k})), \\
&\langle \tilde{b}_{\ell ks} \tilde{b}_{\ell k^\prime s^\prime} \rangle_B \equiv \mathrm{Tr}_B\left[ \tilde{b}_{\ell ks} \tilde{b}_{\ell k^\prime s^\prime} \rho_\ell \right] = \langle \tilde{b}_{\ell  ks}^\dagger \tilde{b}_{\ell k^\prime s^\prime}^\dagger \rangle_B \equiv \mathrm{Tr}_B\left[ \tilde{b}_{\ell ks}^\dagger \tilde{b}_{\ell k^\prime s^\prime}^\dagger \rho_\ell \right]= 0,
\end{aligned}
\end{equation}
where $n_\ell(\tilde{\omega}_{\ell k})$ is the Fermi occupation number defined as $n_\ell(E) = 1/(1+e^{\beta_\ell E})$.

Expanding Eq.~(\ref{LVN}) to the second order in $H_\mathrm{I}$, tracing over the bath degrees of freedom, we obtain
\begin{equation}\label{ME0}
\begin{aligned}
&\dot{\rho}^I(t) = -i \mathrm{Tr_B}\left(\left[ H_I^I(t), \rho_\mathrm{tot}(0) \right]\right) -\int_0^{t} dt' \mathrm{Tr_B}\left(\left[ H_I^I(t), \left[H_I^I(t-t'),\rho_\mathrm{tot}(t-t') \right] \right]\right).
\end{aligned}
\end{equation} 
Using the Born approximation~(\ref{Bornapp}), we first assume that the first term vanishes and then perform the Markov approximation by setting $\rho^I(t-\tau) \approx \rho^I(t)$ and extending the upper limit of integration to infinity. Expanding the double commutator of Eq.~(\ref{ME0}), we obtain
\begin{equation}\label{EQ00}
\begin{aligned}
&\dot{\rho}(t) =
- \sum_{\ell}\sum_{s,s'} \int_0^t \bigg[ 
 \langle B_{\ell s}^\dagger(t) B_{\ell s'}(t-t') \rangle_B c_s(t) c_{s'}^\dagger(t-t') \rho(t)  
+ \langle B_{\ell s}(t) B^\dagger_{\ell s'}(t-t') \rangle_B  c_s^\dagger(t) c_{s'}(t-t') \rho(t) \bigg. \\
&\hspace{3cm}
- \langle B_{\ell s'}(t-t') B^\dagger_{\ell s}(t) \rangle_B c_s(t) \rho(t-t') c_{s'}^\dagger(t-t') 
- \langle B_{\ell s'}^\dagger(t-t') B_{\ell s}(t) \rangle_B c_s^\dagger(t) \rho(t) c_{s'}(t-t') \\[5pt] 
&\hspace{3cm}
- \langle B_{\ell s}(t) B^\dagger_{\ell s'}(t-t') \rangle_B  c_{s'}(t-t') \rho(t) c_s^\dagger(t) 
- \langle B_{\ell s}^\dagger(t) B_{\ell s'}(t-t') \rangle_B c_{s'}^\dagger(t-t') \rho(t) c_s(t) \\
&\hspace{3cm} \bigg.
+ \langle B_{\ell s'}^\dagger(t-t') B_{\ell s}(t) \rangle_B \rho(t) c_{s'}(t-t') c_s^\dagger(t) 
+ \langle B_{\ell s'}(t-t') B^\dagger_{\ell s}(t) \rangle_B  \rho(t) c_{s'}^\dagger(t-t') c_s(t) \bigg] dt' \\
&\quad\quad\quad + 
\sum_{\ell }\sum_{s,s'} \int_0^t \bigg[ 
 \langle B_{\ell s}^\dagger(t) B_{\ell s'}^\dagger(t-t')\rangle_B c_s(t) c_{s'}(t-t') \rho(t)  
+  \langle B_{\ell s}(t) B_{\ell s'}(t-t')\rangle_B c_s^\dagger(t) c_{s'}^\dagger(t-t') \rho(t) \bigg. \\
&\hspace{3cm}
- \langle B_{\ell s'}^\dagger(t-t') B_{\ell s}^\dagger(t)\rangle_B c_s(t) \rho(t) c_{s'}(t-t') 
- \langle B_{\ell s'}(t-t') B_{\ell s}(t)\rangle_B c_s^\dagger(t) \rho(t) c_{s'}^\dagger(t-t') \\[5pt] 
&\hspace{3cm}
- \langle B_{\ell s}^\dagger(t) B_{\ell s'}^\dagger(t-t')\rangle_B c_{s'}(t-t') \rho(t) c_{s}(t) 
- \langle B_{\ell s}(t) B_{\ell s'}(t-t')\rangle_B c_{s'}^\dagger(t-t') \rho(t) c_s^\dagger(t) \\
&\hspace{3cm} \bigg.
+  \langle B_{\ell s'}^\dagger(t-t') B_{\ell s}^\dagger(t)\rangle_B  \rho(t) c_{s'}(t-t') c_s(t) 
+ \langle B_{\ell s'}(t-t') B_{\ell s}(t)\rangle_B \rho(t) c_{s'}^\dagger(t-t') c_s^\dagger(t) \bigg] dt', \\
\end{aligned}
\end{equation}
where we removed the superscript $^I$ to not burden the notation and where we define $B_{\ell s}(t) = \kappa_\ell \sum_{k} b_{\ell ks}(t)$.

\subsection{Bath correlation functions}

The first summation of the right-hand-side of Eq.~(\ref{EQ00}) contains non-vanishing terms only for $s = s'$. They involve bath correlations~(\ref{thermal}) with coefficient $|u_{\ell k}|^2$ or $|v_{\ell k}|^2$. Explicitly, the first term for $s = s' = \downarrow$ involves the correlation
\begin{equation}\label{firstcorr}
\begin{aligned}
\langle B_{\ell \downarrow}^\dagger(t) B_{\ell \downarrow}(t-t') \rangle_B  &= \kappa_\ell^2 \sum_k e^{i V_\ell t'} \left(|u_{\ell k}|^2  e^{i \tilde{\omega}_{\ell k} t'} n(\tilde{\omega}_{\ell k}) + |v_{\ell k}|^2 e^{-i \tilde{\omega}_{\ell k} t'} [1-n(\tilde{\omega}_{\ell k})]\right) 
 \\ 
&= \frac{\gamma_\ell}{\pi} \int_{-\infty}^{\infty} d\omega e^{i V_\ell t'} \left(|u_\ell(\omega)|^2  e^{i \tilde{\omega}_\ell t'} n(\tilde{\omega}_\ell) + |v_\ell(\omega)|^2 e^{-i \tilde{\omega}_\ell t'} [1-n(\tilde{\omega}_\ell)])\right) \\
&= \frac{\gamma_\ell}{\pi}\int_{0}^{\infty} d\omega e^{i V_\ell t'} \left(e^{i \tilde{\omega}_\ell t'} n(\tilde{\omega}_\ell) + e^{-i \tilde{\omega}_\ell t'} [1-n(\tilde{\omega}_\ell)])\right) \\
&= \frac{\gamma_\ell}{\pi}  \int_{0}^{\infty} d\tilde{\omega} e^{i V_\ell t'} D_\ell(\tilde{\omega})\left(e^{i \tilde{\omega} t'} n(\tilde{\omega}) + e^{-i \tilde{\omega} t'} [1-n(\tilde{\omega})])\right) \\
&= \frac{\gamma_\ell}{\pi} \int_{-\infty}^{\infty} d\tilde{\omega} D_\ell(\tilde{\omega}) e^{i (\tilde{\omega} + V_\ell) t'} n(\tilde{\omega})
 \end{aligned}
\end{equation}
In Eq.~(\ref{firstcorr}), $\gamma_\ell = \pi \kappa_\ell^2 \rho_N$ is the tunnelling rate between the QD and the lead $\ell$, where $\rho_N = \mathcal{V} m k_{\ell F}/2\pi^2 \hbar^2$ is the density of states of a normal lead assumed to be constant around the Fermi level, where $\mathcal{V}$ is the volume of the lead $\ell$, $m$ the electron mass and $k_{\ell F}$ the Fermi wavenumber, as in~\cite{Pfa13}; and
\begin{equation}\label{DOS}
D_\ell(E) = \frac{|E|}{\sqrt{E^2 - \Delta_\ell^2}} \Theta( |E| - \Delta_\ell)
\end{equation}
is the adimensional density of states of a superconducting lead, which naturally appears when making the substitution $\omega \to \tilde{\omega} = \sqrt{\omega^2 + \Delta_\ell^2}$ in the integration. A quick look shows that the right-hand-side of Eq.~(\ref{EQ00}) contains only the following two distinct correlations functions
\begin{equation}\label{fpm}
\begin{aligned}
f_{\ell +}(t') &= \langle B_{\ell s}^\dagger(t) B_{\ell s}(t-t') \rangle_B = \frac{\gamma_\ell}{\pi}  \int_{-\infty}^{\infty} d\tilde{\omega} D_\ell(\tilde{\omega}) e^{i (\tilde{\omega} + V_\ell) t'} n(\tilde{\omega}),\\
f_{\ell -}(t') &= \langle B_{\ell s}(t) B_{\ell s}^\dagger(t-t') \rangle_B = \frac{\gamma_\ell}{\pi}  \int_{-\infty}^{\infty} d\tilde{\omega} D_\ell(\tilde{\omega}) e^{-i (\tilde{\omega} + V_\ell) t'} [1 - n(\tilde{\omega})],
\end{aligned}
\end{equation}
All the other correlations functions are up to a minus sign these correlations functions or their complex conjugates.

The second summation of the right-hand-side of Eq.~(\ref{EQ00}) contains non-vanishing terms only for $s \neq s'$. They also involve bath correlations~(\ref{thermal}) but with coefficients $u v$ or $u_{\ell k}^*v_{\ell k}^*$. There are only two distinct correlations functions $f_{\ell 1}(t')$ and $f_{\ell 2}(t')$ (all the others are up to a minus sign these correlations functions or their conjugates) which read
\begin{equation}\label{f12}
\begin{aligned}
f_{\ell 1}(t-t') &= \langle B_{\ell \downarrow}^\dagger(t) B_{\ell \uparrow}^\dagger(t-t') \rangle_B = \frac{2\gamma_\ell}{\pi}  \, e^{2i V_\ell t} \int_0^\infty d\tilde{\omega} \, D_\ell(\tilde{\omega}) u_\ell^*(\tilde{\omega}) v_\ell^*(\tilde{\omega}) e^{-i V_\ell t'} \left(-  e^{i \tilde{\omega} t'} n(\tilde{\omega}) + e^{-i \tilde{\omega} t'} [1-n(\tilde{\omega})] \right), \\
f_{\ell 2}(t-t') &= \langle B_{\ell \uparrow}^\dagger(t-t') B_{\ell \downarrow}^\dagger(t) \rangle_B = \frac{2\gamma_\ell}{\pi}  \, e^{2i V_\ell t}  \, \int_0^\infty d\tilde{\omega} \, D_\ell(\tilde{\omega}) u_\ell^*(\tilde{\omega}) v_\ell^*(\tilde{\omega}) e^{-i V_\ell t'} \left(  e^{-i \tilde{\omega} t'} n(\tilde{\omega}) - e^{i \tilde{\omega} t'} [1-n(\tilde{\omega})] \right).
\end{aligned}
\end{equation}
Note that while the correlations~(\ref{fpm}) depend only on the time difference $t'$, Eq.~(\ref{f12}) explicitly depend on time $t$.

\subsection{Case $g_\ell = 0$ - Standard Born-Markov master equation}

Before deriving the general Floquet-Born-Markov equation for the QD, we first investigate the case of $g_\ell = 0$ for which the QD Hamiltonian is time-independent. This corresponds to the case of a standard two-terminal junction, where the QD is only coupled to two superconducting leads. This yields a standard Born-Markov master equation, similar to the one derived in~\cite{Kos13, Pfa13}. 

The QD Hamiltonian $H_\mathrm{QD} = \sum_{s = \downarrow, \uparrow} \omega c_s^\dagger c_s + U c_\uparrow^\dagger c_\uparrow c_\downarrow^\dagger c_\downarrow$ can be diagonalised in the basis $\left\{\ket{0},\ket{\downarrow},\ket{\uparrow}, \ket{\downarrow\uparrow} \right\}$ by writing the operators $c_s$ in terms of the projection operators $\sigma_{ij} = \ket{i}\bra{j}$ ($i,j = 0, \downarrow,\uparrow,\downarrow\uparrow$) satisfying $\sigma_{ij}\sigma_{kl} = \delta_{jk}\sigma_{il}$, 
\begin{equation}\label{cs}
\begin{aligned}
&c_\downarrow = \ket{0}\bra{\downarrow} + \ket{\uparrow}\bra{\downarrow \uparrow}, \\
&c_\uparrow = \ket{0}\bra{\uparrow} - \ket{\downarrow}\bra{\downarrow \uparrow}, \\
\end{aligned}
\end{equation}
as in~\cite{Zha11}. Note that we chose the convention $\ket{\downarrow\uparrow} = c_\downarrow^\dagger c_\uparrow^\dagger \ket{0}$ instead of $\ket{\downarrow\uparrow} = c_\uparrow^\dagger c_\downarrow^\dagger \ket{0}$ regarding the order of action of creation operators to find their matrix representation (otherwise the sign in front of $\ket{s}\bra{\downarrow \uparrow}$ would be reversed). The system operators in interaction picture with respect to $H_0$ then simply reads
\begin{equation}\label{intcs}
\begin{aligned}
&c_\downarrow(t) = \ket{0}\bra{\downarrow} e^{-i \omega t} + \ket{\uparrow}\bra{\downarrow\uparrow} e^{-i (U+\omega)t}, \\
&c_\uparrow(t) = \ket{0}\bra{\uparrow} e^{-i \omega t} - \ket{\downarrow}\bra{\downarrow\uparrow} e^{-i (U+\omega)t}. \\
\end{aligned}
\end{equation}

From Eq.~(\ref{EQ00}), using Eq~(\ref{intcs}) and the expressions of the bath correlation functions~(\ref{fpm}) and (\ref{f12}), we obtain the master equation in the standard form (coming back to the Schrödinger picture)
\begin{equation}\label{MEmodel1}
\boxed{
\dot{\rho} = -i \left[ H_{\mathrm{QD}} + H_{\mathrm{LS}}^{(s)} + H_{\mathrm{LS}}^{(p)}(t), \rho \right] + \mathcal{D}^{(s)}\left(\rho\right) + \mathcal{D}^{(p)}\left(\rho,t\right)
}
\end{equation}
where
\begin{equation}\label{HLSs}
H_{\mathrm{LS}}^{(s)} = \sum_{\ell ,s} \left(  \Omega_{\ell +}(-\omega + V_\ell ) \sigma_{0s}\sigma_{0s}^\dagger + \Omega_{\ell -}(\omega - V_\ell ) \sigma_{0s}^\dagger\sigma_{0s} + \Omega_{\ell +}(-U-\omega + V_\ell ) \sigma_{s\downarrow\uparrow}\sigma_{s \downarrow\uparrow}^\dagger + \Omega_{\ell -}(U+\omega - V_\ell ) \sigma_{s\downarrow\uparrow}^\dagger\sigma_{s\downarrow\uparrow}\right) 
\end{equation}
is the Lamb-shift Hamiltonian coming from standard single particle tunnelling processes, where
\begin{equation}\label{HLSp}
\begin{aligned}
H_{\mathrm{LS}}^{(p)}(t) &= i \sum_\ell  \Big(- \Big[\Gamma_{\ell 1}(U+\omega - V_\ell,t) + \Gamma_{\ell 2}(\omega - V_\ell ,t)\Big] \sigma_{0\downarrow\uparrow}
+ \Big[\Gamma_{\ell 1}^*(U+\omega - V_\ell ,t) + \Gamma_{\ell 2}^*(\omega - V_\ell,t )\Big] \sigma_{\downarrow\uparrow 0} \Big)
\end{aligned} 
\end{equation}
is a time-dependent Lamb-shift Hamiltonian coming from Cooper-pair tunnelling processes, where
\begin{equation}\label{Ds}
\begin{aligned}
\mathcal{D}^{(s)}\left(\rho\right)  &=  2 \sum_{\ell ,s} \left(
\gamma_{\ell -}(\omega - V_\ell ) \left( \sigma_{0s} \rho \sigma_{0s}^\dagger - \frac{1}{2} \left\{ \sigma_{0s}^\dagger\sigma_{0s}, \rho \right\} \right) 
+ \gamma_{\ell -}(U + \omega - V_\ell ) \left( \sigma_{\overline{s}\downarrow\uparrow} \rho \sigma_{\overline{s}\downarrow\uparrow}^\dagger - \frac{1}{2} \left\{ \sigma_{s\downarrow\uparrow}^\dagger\sigma_{s\downarrow\uparrow}, \rho \right\} \right) \right.  \\
& \hspace{1.3cm}  \left.
+  \gamma_{\ell +}(-\omega + V_\ell ) \left( \sigma_{0s}^\dagger \rho \sigma_{0s} - \frac{1}{2} \left\{ \sigma_{0s}\sigma_{0s}^\dagger, \rho \right\} \right) + \gamma_{\ell +}(-U - \omega + V_\ell ) \left( \sigma_{s\downarrow\uparrow}^\dagger \rho \sigma_{s\downarrow\uparrow} - \frac{1}{2} \left\{ \sigma_{s\downarrow\uparrow}\sigma_{s\downarrow\uparrow}^\dagger, \rho \right\} \right)  \right)
\end{aligned}
\end{equation}
is a time-independent single-particle dissipator coming from standard single-particle tunnelling processes; and where
\begin{equation}\label{Dp}
\begin{aligned}
\mathcal{D}^{(p)}\left(\rho,t\right) &= \sum_{\ell ,s} \left( \Big[ \Gamma_{\ell 2}(U+\omega - V_\ell,t ) - \Gamma_{\ell 1}(\omega - V_\ell,t )\Big] \sigma_{0s} \rho \sigma_{s\downarrow\uparrow} \right.  +  \Big[ \Gamma_{\ell 2}^*(U+\omega - V_\ell,t ) - \Gamma_{\ell 1}^*(\omega - V_\ell ,t ) \Big] \sigma_{s\downarrow\uparrow}^\dagger \rho \sigma_{0s}^\dagger \\
& \hspace{0.5cm} +  \Big[ \Gamma_{\ell 1}^*(U+\omega - V_\ell ,t) - \Gamma_{\ell 2}^*(\omega - V_\ell,t ) \Big] \left( \sigma_{0s}^\dagger \rho \sigma_{s\downarrow\uparrow}^\dagger  - \frac{1}{2} \left\{ \sigma_{s\downarrow\uparrow}^\dagger \sigma_{0s}^\dagger , \rho \right\} \right)\\ 
& \hspace{0.5cm} + \bigg. \Big[ \Gamma_{\ell 1}(U+\omega - V_\ell,t ) - \Gamma_{\ell 2}(\omega - V_\ell,t )\Big] \left( \sigma_{s\downarrow\uparrow} \rho \sigma_{0s} - \frac{1}{2} \left\{ \sigma_{0s}\sigma_{s\downarrow\uparrow}, \rho \right\} \right) \bigg)
\end{aligned}
\end{equation}
is a time-dependent Cooper-pair dissipator coming from Cooper-pair tunnelling processes. Note that the two first terms on the right-hand side do not have the anticommutator since $\sigma_{s\downarrow\uparrow}\sigma_{0s} = 0$ and $\sigma_{0s}^\dagger\sigma_{s\downarrow\uparrow}^\dagger = 0$.

The master equation involves the complex rates
\begin{equation}\label{complexrates}
\begin{aligned}
\Gamma_{\ell\pm}(E \pm V_\ell ) &= \int_0^\infty dt' f_{\ell \pm}(t') e^{i E t'} = \gamma_{\ell \pm}(E  \pm V_\ell ) + i \Omega_{\ell \pm}(E  \pm V_\ell ), \\
\Gamma_{\ell j}(E - V_\ell ,t) &= \int_0^\infty dt' f_{\ell \alpha}(t-t') e^{i E t'} = e^{2i V_\ell t} \big[ \gamma_{\ell \alpha}(E - V_\ell) + i \Omega_{\ell \alpha}(E  - V_\ell) \big],
\end{aligned}
\end{equation}
where $j = 1,2$, where $E$ corresponds to system transition energies and where $\gamma_{\ell \pm}$ ($\gamma_{\ell j}$) and $\Omega_{\ell \pm}$ ($\Omega_{\ell j}$) are the real and imaginary parts of $\Gamma_{\ell \pm}$ ($\Gamma_{\ell j} e^{-2i V_\ell t}$) which explicitly read
\begin{equation}\label{rates}
\begin{aligned}
&\gamma_{\ell \pm}(E) = \gamma_\ell D_\ell(E) [1-n(E)], \\
&\gamma_{\ell 1}(E) = -\gamma_{\ell 2}(-E) = 2 \gamma_\ell  D_\ell(E) u_\ell^*(E)v_\ell^*(E) [1 - n(E)] \left[\Theta(E) - \Theta(-E)\right], \\
&\Omega_{\ell \pm}(E) =  \frac{\gamma_\ell}{\pi} \,\mathrm{P.V.} \int_{-\infty}^{\infty} d\tilde{\omega}D_\ell(\tilde{\omega}) \frac{n(\tilde{\omega})}{E + \tilde{\omega}}, \\
&\Omega_{\ell 1}(E) = \Omega_{\ell 2}(-E) = \frac{2\gamma_\ell}{\pi} \,\mathrm{P.V.} \int_{0}^{\infty} d\tilde{\omega}D_\ell(\tilde{\omega}) u_{\ell}^{*}(\tilde{\omega})v_{\ell}^{*}(\tilde{\omega}) \left( -\frac{n(\tilde{\omega})}{E + \tilde{\omega}}  + \frac{1-n(\tilde{\omega})}{E - \tilde{\omega}}  \right),
\end{aligned}
\end{equation}
where $\mathrm{P.V.\ }$ denotes the principal value.
Note that the integrands appearing in the expressions of the shifts $\Omega_{\ell \alpha}$ ($\alpha = \pm, 1,2$) do not converge for $\omega \to \pm\infty$, and one has to introduce a cutoff frequency $\omega_c$ in the integration domain to obtain finite values for the shifts. This cutoff represents the fact that the bandwidths of real leads are finite.

Writing down explicitly the master equation for the density matrix elements $\rho_{ij} = \bra{ i } \rho \ket{j}$, one can see that the populations $\rho_{00}, \rho_{\downarrow \downarrow}, \rho_{\uparrow \uparrow}, \rho_{\downarrow\uparrow \downarrow\uparrow}$ are only coupled to the coherences $\rho_{0\downarrow\uparrow}$ and $\rho_{\downarrow\uparrow0}$ through the Lamb-shift Hamiltonian $H_{\mathrm{LS}}^{(p)}(t)$ and dissipator $\mathcal{D}^{(p)}\left(\rho,t\right)$. This corresponds to a proximity effect which vanishes for $\Delta_\ell = 0$. Note that in~\cite{Pfa13}, these terms are not present due to the use of a number-conserving Bogoliubov transformation. In~\cite{Kos13}, they are present (for $U + 2 \omega = 0$) but are \textit{time-independent}, because the authors~\cite{Kos13} included the bias voltage by making the replacement $\omega \to \omega -V_\ell$ in the end of the calculations, instead of including it directly in the lead Hamiltonian~(\ref{model2}). We note however that these terms have an important effect only in the very low bias regime $V_\ell \ll \gamma_\ell$. For $V_\ell \gg \gamma_\ell$, one can invoke a secular approximation to neglect them. The master equation then becomes similar to the one derived by Kosov \textit{et al}~\cite{Kos13}: no subgap current can be observed. 

\subsubsection{Large gap limit}\label{LargeGap}
In the large gap limit $\Delta_\ell \gg |\omega \pm V_\ell|, |\omega + U \pm V_\ell|$, all the $\gamma_{\ell \alpha}(E)$ ($\alpha = \pm, 1,2$) vanish since all the transition energies $E$ lie inside the gap. Also, $\Omega_{\ell j}(E)$ becomes independent of $E$, i.e.\ $\Omega_{\ell j}(E) \approx \tilde{\Omega}_{\ell}$, with
\begin{equation}
\tilde{\Omega}_{\ell} = - \frac{\gamma_\ell}{\pi} \Delta_\ell e^{i \phi_\ell} \,\mathrm{P.V.}\int_{0}^{\infty} d\tilde{\omega}\frac{D(\tilde{\omega})}{\tilde{\omega}^2} \Big(n(\tilde{\omega}) + n(-\tilde{\omega})\Big)
\end{equation}
As a consequence, we have
\begin{equation}
\begin{aligned}
&\Gamma_{\ell 2}^*(\omega - V_\ell,t ) - \Gamma_{\ell 1}^*(U+\omega - V_\ell ,t) = 0, \\
&\Gamma_{\ell 1}(\omega - V_\ell,t ) - \Gamma_{\ell 2}(U+\omega - V_\ell,t ) = 0,
\end{aligned}
\end{equation}
which implies that $\mathcal{D}^{(p)}\left(\rho,t\right) = 0$; and
\begin{equation}\label{DrivingH}
\boxed{
H_{\mathrm{LS}}^{(p)}(t) \approx \sum_\ell  \Big( g_\ell c_\downarrow c_\uparrow e^{2i V_\ell  t} + \mathrm{h.c.} \Big)
}
\end{equation}
where $g_\ell = -2\tilde{\Omega}_\ell$. Overall, the master equation reduces to
\begin{equation}
\dot{\rho} = -i \left[ H_{\mathrm{QD}} + H_{\mathrm{LS}}^{(s)} + H_{\mathrm{LS}}^{(p)}(t), \rho \right].
\end{equation}
Hence, in the large gap limit, the coupling of the QD to the superconductors reduces to a driving of the transition between the non-occupied and double-occupied states of the QD, as it can be derived using other methods (see~\cite{Mar11} and references therein). This justifies the use of an Hamiltonian of the form~(\ref{DrivingH}) in our model~(\ref{model2}) to treat the coupling of the large-gap superconducting leads to the QD.

\subsection{Case $g_\ell \neq 0$ - Floquet-Born-Markov master equation}

We now come back to the general case of $g_\ell \neq 0$ (the four-terminal junction), where the QD Hamiltonian is time-dependent. In order to perform the time-integration in Eq.~(\ref{EQ00}), we evaluate the time dependence of the system operators $c_s(t)$ given by Eq.~(\ref{ctbt}) using the Floquet theory~\cite{Gri98}. For that purpose, we suppose in the following that $V_L = -V_R = V/2$, so that the effective QD Hamiltonian is periodic of period $T = 2\pi/V$. If it was not the case, one could simply work in the rotating-frame with respect to one of the driving frequency $2V_\ell$, let say $2V_L$. This would provide an periodic Hamiltonian of period $\delta = 2(V_R - V_L)$, and the same theory would apply.

Since the effective QD Hamiltonian is periodic, the QD wavefunction $\ket{\psi(t)}$ satisfying the Schr\"odinger equation
\begin{equation}\label{TDse}
i \frac{d}{dt}\ket{\psi(t)} = H_\mathrm{QD}^\mathrm{eff}(t) \ket{\psi(t)}
\end{equation}
can be written as
\begin{equation}
\ket{\psi(t)} = \sum_a d_a \ket{\psi_a(t)} = \sum_a d_a \, e^{- i E_a t}\ket{\phi_a(t)},
\end{equation}
where $\ket{\psi_a(t)} = e^{- i E_a t}\ket{\phi_a(t)}$ are the Floquet states with the \textit{periodic} Floquet modes $\ket{\phi_a(t+T)} = \ket{\phi_a(t)}$, quasi-energies $E_a$, and $d_a = \langle \phi_a(0) | \psi(0) \rangle$. By definition of the propagator~(\ref{propa}), we have
\begin{equation}\label{EVeq}
\ket{\psi_a(T)} = U(T) \ket{\psi_a(0)} 
\Leftrightarrow e^{- i E_a T}\ket{\phi_a(0)} = U(T) \ket{\phi_a(0)},
\end{equation}
showing that $e^{- i E_a T}$ are the eigenvalues of $U(T)$, which can be numerically computed using $U(T) \approx \prod_{n = 0}^{N} e^{ - i H_\mathrm{QD}^\mathrm{eff}(n dt) dt}$ with $N = T/dt - 1$. Solving the eigenvalue problem~(\ref{EVeq}), we obtain $E_{a,k} = E_a + k \frac{2\pi}{T}$ with $k \in \mathbb{Z}$, and consider the values of $E_{a,k}$ lying in the first Brillouin zone $[-\pi/T, \pi/T]$ to define the quasienergies $E_a$. The eigenvectors correspond to the Floquet modes at initial time $\ket{\phi_a(0)}$. The Floquet modes at all times $t$ are obtained from these latters using
\begin{equation}
\ket{\phi_a(t)} = e^{i E_a t}U(t)\ket{\phi_a(0)}.
\end{equation}

\subsubsection{Master equation in the Floquet basis}

We now decompose the density matrix in the Floquet mode basis $\{\ket{\phi_a(0)}\}$, i.e.\ 
\begin{equation}\label{rhoI}
\rho^I(t) = \sum_{a,b} \rho^{I,ab}(t) \ket{\phi_a(0)} \bra{\phi_b(0)},
\end{equation}
where we restored for clarity the label $^I$ denoting the interaction picture for the density matrix.
We derive below the equations of motion for the density matrix element $\rho^{I,ab}(t) \equiv \bra{\phi_{a}(0)}\rho^I(t)\ket{\phi_b(0)}$ from Eq.~(\ref{EQ00}).

In this basis, the matrix elements of the system operator $c_s(t)$ reads
\begin{equation}
\bra{\phi_{a}(0)}c_s(t)\ket{\phi_b(0)} = \bra{\phi_{a}(t)}c_s\ket{\phi_b(t)} e^{i (E_a - E_b) t}.
\end{equation}
Since $\ket{\phi_a(t)}$ is periodic of period $T$, we can rewrite $\bra{\phi_{a}(t)}c_s\ket{\phi_b(t)}$
in the Fourier space as
\begin{equation}\label{fourierseries}
\bra{\phi_{a}(t)}c_s\ket{\phi_b(t)} = \sum_{k\in \mathbb{Z}} e^{i k V t} c_s^{ab k},
\end{equation}
which yields
\begin{equation}
\bra{\phi_{a}(t)}c_s\ket{\phi_b(t)} e^{i (E_a - E_b) t} = \sum_{k\in \mathbb{Z}} e^{i k \Delta_{ab k} t} c_s^{ab k} ,
\end{equation}
where $\Delta_{a b k} = E_a - E_b + k V$ and
\begin{equation}
c_s^{ab k} = \frac{1}{T} \int_0^T e^{-i k V t} \bra{\phi_{a}(t)}c_s\ket{\phi_b(t)} dt.
\end{equation}
Using the Floquet basis, the first term of the right-hand side of the master equation~(\ref{EQ00}) reads
\begin{equation}
\begin{aligned}
\langle\phi_a(0)| & \left( \int_0^\infty dt' f_{\ell +}(t')  c_s(t)c_s^\dagger(t-t') \rho(t) \right) |\phi_b(0)\rangle 
= \sum_{c,d}\sum_{k,k'} e^{i (\Delta_{ack} + \Delta_{cdk'}) t} c_s^{ack} c_s^{\dagger cdk'} \rho^{I,db}(t) \int_0^\infty  dt' f_{\ell +}(t')  e^{-i \Delta_{cdk'} t'},
\end{aligned}
\end{equation}
and all other terms can be written in the same way. Hence, we see that the master equation involves the complex rates~(\ref{complexrates}). 
All together, the master equation~(\ref{EQ00}) written in the Floquet basis gives us the following set of equations for the matrix elements  $\rho^{I,ab}(t) \equiv \bra{\phi_{a}(0)}\rho^I(t)\ket{\phi_b(0)}$, i.e.\ the non-secular Floquet-Born-Markov master equation
\begin{equation}\label{EQF}\boxed{
\dot{\rho}^{I,ab}(t) = \sum_\ell (\mathcal{L}_\ell^{(s)}[\rho^I(t)])^{ab}+ \sum_\ell (\mathcal{L}_\ell^{(p)}[\rho^I(t)])^{ab}}
\end{equation}
where
\begin{equation}
\begin{aligned} 
(\mathcal{L}_\ell^{(s)}[\rho^I(t)])^{ab} &= - \sum_{s} \sum_{k,k'} \sum_{c,d} \left[ e^{i (\Delta_{cdk'} + \Delta_{ack})t}
 \left( c_s^{ack} c_s^{\dagger cdk'}  \Gamma_{\ell +}(-\Delta_{cdk'}+V_\ell)  +
 c_s^{\dagger ack} c_s^{cdk'} \Gamma_{\ell -}(-\Delta_{cdk'}-V_\ell)\right) \rho^{I,db}(t) \right. \\
&\hspace{1cm}\phantom{\times\bigg[}
- e^{i (\Delta_{dbk'} + \Delta_{ack})t} \left( c_s^{ack} c_s^{\dagger dbk'} \Gamma_{\ell -}^*(\Delta_{dbk'}-V_\ell) + c_s^{\dagger ack} c_s^{dbk'} \Gamma_{\ell +}^*(\Delta_{dbk'}+V_\ell)\right) \rho^{I,cd}(t) \\
&\hspace{1cm}\phantom{\times\bigg[}
- e^{i (\Delta_{dbk'} + \Delta_{ack})t} \left(  c_s^{ack} c_s^{\dagger dbk'}  \Gamma_{\ell -}(-\Delta_{ack}-V_\ell) + c_s^{\dagger ack} c_s^{dbk'} \Gamma_{\ell +}(-\Delta_{ack}+V_\ell)\right) \rho^{I,cd}(t)\\
&\hspace{1cm}\phantom{\times\bigg[}  +
\left.   e^{i (\Delta_{dbk'} + \Delta_{cdk})t} \left( c_s^{cdk} c_s^{\dagger dbk'} \Gamma_{\ell +}^*(\Delta_{cdk}+V_\ell) + c_s^{\dagger cdk} c_s^{dbk'} \Gamma_{\ell -}^*(\Delta_{cdk}-V_\ell)\right) \rho^{I,ac}(t) \right]\\
\end{aligned}
\end{equation}
is the Linbladian coming from standard single-particle tunnelling processes, and where
\begin{equation}\label{MEp}
\begin{aligned}
(\mathcal{L}_\ell^{(p)}[\rho^I(t)])^{ab} &= \sum_{c,d}\sum_{k,k'} \left[ 2 e^{i (\Delta_{cdk'} + \Delta_{ack})t}
 \left(  c_\downarrow^{ack} c_\uparrow^{cdk'}  \Gamma_{\ell 1}(-\Delta_{cdk'}-V_\ell,t) +  c_\downarrow^{\dagger ack} c_\uparrow^{\dagger cdk'}  \Gamma_{\ell 2}^*(\Delta_{cdk'}-V_\ell,t)\right) \rho^{I,db}(t) \right. \\
&+
 e^{i (\Delta_{dbk'} + \Delta_{ack})t} \left( c_\uparrow^{ack} c_\downarrow^{dbk'}  \Gamma_{\ell 2}(-\Delta_{dbk'}-V_\ell,t)- c_\downarrow^{ack} c_\uparrow^{dbk'}  \Gamma_{\ell 2}(-\Delta_{dbk'}-V_\ell,t)\right) \rho^{I,cd}(t) \\
&+ e^{i (\Delta_{dbk'} + \Delta_{ack})t}
\left(  c_\uparrow^{\dagger ack} c_\downarrow^{\dagger dbk'}  \Gamma_{\ell 1}^*(\Delta_{dbk'}-V_\ell,t) - c_\downarrow^{\dagger ack} c_\uparrow^{\dagger dbk'}  \Gamma_{\ell 1}^*(\Delta_{dbk'}-V_\ell,t)\right) \rho^{I,cd}(t) \\
&- e^{i (\Delta_{dbk'} + \Delta_{ack})t} \left( c_\uparrow^{ack} c_\downarrow^{dbk'}  \Gamma_{\ell 1}(-\Delta_{ack}-V_\ell,t)- c_\downarrow^{ack} c_\uparrow^{dbk'}  \Gamma_{\ell 1}(-\Delta_{ack}-V_\ell,t)\right) \rho^{I,cd}(t) \\
&- e^{i (\Delta_{dbk'} + \Delta_{ack})t} \left( c_\uparrow^{\dagger ack} c_\downarrow^{\dagger dbk'}  \Gamma_{\ell 2}^*(\Delta_{ack}-V_\ell,t) - c_\downarrow^{\dagger ack} c_\uparrow^{\dagger dbk'}  \Gamma_{\ell 2}^*(\Delta_{ack}-V_\ell,t)\right) \rho^{I,cd}(t) \\
&+ \left. 2 e^{i (\Delta_{cdk} + \Delta_{dbk'})t} \left( c_\uparrow^{cdk} c_\downarrow^{dbk'}  \Gamma_{\ell 2}(-\Delta_{cdk}-V_\ell,t) + c_\uparrow^{\dagger cdk} c_\downarrow^{\dagger dbk'}  \Gamma_{\ell 1}^*(\Delta_{cdk}-V_\ell,t)\right) \rho^{I,ac}(t) \right]
\end{aligned}
\end{equation}
is the Linbladian coming from Cooper-pair tunnelling processes.



The master equation for the matrix elements in Schrödinger picture $\rho^{ab}(t)$ can be obtained from Eq.~(\ref{EQF}) by making the replacement $\rho^{I,ab}(t) = e^{i (E_a - E_b)t} \rho^{ab}(t)$. In doing so, one can see that all terms $e^{i(\Delta_{cdk'} + \Delta_{ack})t}$ reduces to $e^{i(k + k')Vt}$, showing that the master equation exhibits the same periodicity than the effective QD Hamiltonian. This implies that the steady state of the master equation is also periodic with the same period $T = 2\pi/V$~\cite{Yud16}.

\section{Solution of the master equation and particle currents}

The master equation~(\ref{EQF}) can be solved using different methods: either via brute force methods without any further treatment or by first writing Eq.~(\ref{EQF}) in Schrödinger picture and then exploiting its periodicity. One can indeed vectorize the density matrix as $\ket{\rho^S(t)} = (\rho^{S,11}(t),\rho^{S,12}(t),\dotsc, \rho^{S,44}(t))$ to obtain the equation of motion
\begin{equation}\label{vecME}
\ket{\dot{\rho}^S(t)} = L(t)  \ket{\rho^S(t)}
\end{equation}
where $L(t)$ is a periodic time-dependent matrix of period $T$, and then apply again the Floquet theory on Eq.~(\ref{vecME})~\cite{Yud16}. Another solution consists in targeting directly the periodic steady state of Eq.~(\ref{vecME}) by writing it in Fourier space and solving the linear set of equations for its Fourier components~\cite{Mal16}. In the main text, we solved the master equation~(\ref{EQF}) via brute force, but compared the solutions with these other methods and obtained the same results.

From the solution of the master equation~(\ref{EQF}) for the matrix elements $\rho^{I,ab}(t)$, one can rebuild the whole density matrix~(\ref{rhoI}) and the expectation values of any system operator $O$ through
\begin{equation}\label{expop}
\begin{aligned}
\langle O \rangle &= \mathrm{Tr}\left[ U^\dagger(t) O U(t) \rho^I(t) \right] = \sum_{j = 0,\downarrow,\uparrow,\downarrow \uparrow}\sum_{ab} \rho^{I,ab}(t) e^{-i (E_a - E_b) t} \bra j O \ket{\phi_a(t)} \bra{\phi_b(t)} j \rangle.
\end{aligned}
\end{equation}
We derive now the expressions of the particle currents $I_\mathrm{QD}$, $I_\ell^{(s)}$ and $I_\ell^{(p)}$ respectively in the QD, the moderate-gap and the large-gap superconducting leads $\ell$. The choice of the symbols $(s)$ and $(p)$ expresses the fact that the current in the moderate-gap superconducting leads is mainly due to single-particle ($s$) tunnelling, while the current in the large-gap superconducting leads is only due to pair ($p$) tunnelling. They are defined as
\begin{align}
&I_\mathrm{QD} = \frac{d}{dt}\Big( \sum_{s} c_{ s}^\dagger c_{s} \Big) = -i \kappa_\ell \sum_{ks} \left( c_s^\dagger b_{\ell k s } - \mathrm{h.c.}\right) \nonumber - 2 i   \sum_\ell \Big(  g_\ell^* e^{-2i\mu_F t} c_\uparrow^\dagger c_\downarrow^\dagger - \mathrm{h.c.} \Big), \\
&I_\ell^{(s)} = \frac{d}{dt}\Big( \sum_{ks} b_{\ell k s}^\dagger b_{\ell k s} \Big) = i \kappa_\ell \sum_{ks} \left( c_s^\dagger b_{\ell k s } - \mathrm{h.c.}\right), \\
&I_\ell^{(p)} = \frac{d}{dt}\left( S_{\ell}^\dagger S_{\ell} \right) = i \left( g_\ell^* e^{-2i V_\ell t} c_\uparrow^\dagger c_\downarrow^\dagger - \mathrm{h.c.} \right), \label{Ip} \\
\end{align}
where $S_\ell$ are the Cooper-pair annihilation operators introduced in Eq.~(\ref{CPtunnelling}) and where we used the Langevin equations for $c_s$, $b_{\ell k s}$ and $S_\ell$, i.e.\
\begin{equation}
\begin{aligned}
&\dot{c}_s =  i \left[ H_\mathrm{QD}, c_s\right] - i \sum_\ell \kappa_\ell \sum_{k} b_{\ell k s} + i \sum_\ell g_\ell^* S_\ell \left( c_\uparrow^\dagger \left\{ c_\downarrow^\dagger, c_s\right\} - \left\{ c_\uparrow^\dagger, c_s \right\} c_\downarrow^\dagger \right) \\
&\dot{b}_{\ell k s} = i \left[ H_\mathrm{lead}, b_{\ell ks} \right] - i \kappa_\ell c_s, \\
&\dot{S}_{\ell} = -2 i V_\ell S_\ell  - i g_\ell c_\downarrow c_\uparrow .
\end{aligned}
\end{equation}
Note that the total number of particles is conserved, i.e.,
\begin{equation}\label{continuity}
I_\mathrm{QD} + \sum_{\ell} I_\ell^{(s)} + 2 \sum_\ell I^{(p)}_\ell = 0,
\end{equation}
where the factor $2$ is front of $I_\ell^{(p)}$ denotes the fact that a Cooper-pair is made of two electrons.

The expectation value of the particles current in the QD is obtained from the solutions of the master equation and Eq.~(\ref{expop}), that is
\begin{equation}
\begin{aligned}
\langle I_\mathrm{QD} \rangle = \sum_s \frac{d}{dt}\langle c_s^\dagger c_s \rangle  &= \sum_s \frac{d}{dt}\mathrm{Tr}\left[U^\dagger(t) c_s^\dagger c_s U(t) \rho^I(t) \right].
\end{aligned}
\end{equation}
Due to the conservation of particles~(\ref{continuity}), it can be related to the particles current in the leads, as shown below. Applying the derivative and using the fact that $dU(t)/dt = -i H_\mathrm{QD}^\mathrm{eff}(t) U(t)$, we get
\begin{equation}\label{IQDtemp}
\begin{aligned}
\langle I_\mathrm{QD} \rangle
&=   i \sum_s \mathrm{Tr}\left[ U^\dagger(t) H_\mathrm{QD}^\mathrm{eff}(t) c_s^\dagger c_s U(t)\rho^I(t) \right]  - i  \sum_s \mathrm{Tr}\left[ U^\dagger(t) c_s^\dagger c_s H_\mathrm{QD}^\mathrm{eff}(t) U(t) \rho^I(t)  \right] + \sum_s \mathrm{Tr}\left[ U^\dagger(t) c_s^\dagger c_s U(t) \frac{d\rho^I(t)}{dt}  \right]. 
\end{aligned}
\end{equation}
The two first terms on the right-hand-side can be rewritten as
\begin{equation}
\begin{aligned}
&i \sum_s \langle H_\mathrm{QD}^\mathrm{eff}(t) c_s^\dagger c_s \rangle  - i  \sum_s \langle c_s^\dagger c_s  H_\mathrm{QD}^\mathrm{eff}(t) \rangle = 2 i \sum_\ell g_\ell e^{2i V_\ell t} \langle c_\downarrow c_\uparrow \rangle - 2 i \sum_\ell g_\ell^* e^{-2i V_\ell t} \langle c_\uparrow^\dagger c_\downarrow^\dagger \rangle = - 2 \sum_\ell \langle I^{(p)}_\ell \rangle
\end{aligned}
\end{equation}
where [see Eq.~(\ref{Ip})]
\begin{equation}\label{Ipl}
\langle I^{(p)}_\ell \rangle = i \left( g_\ell^* e^{-2i V_\ell t} \langle c_\uparrow^\dagger c_\downarrow^\dagger \rangle - \mathrm{h.c.} \right).
\end{equation}
Finally, replacing the derivative in Eq.~(\ref{IQDtemp}) by the right-hand-side of the master equation~(\ref{EQF}) yields
\begin{equation}
\begin{aligned}
\langle I_\mathrm{QD} \rangle 
&= - 2 \sum_\ell \langle I^{(p)}_\ell \rangle +  \sum_{\ell s} \mathrm{Tr}\left[ U^\dagger(t) c_s^\dagger c_s U(t) \left(\mathcal{L}_\ell \left[  \rho(t)\right]\right) \right] = - 2 \sum_\ell  \langle I^{(p)}_\ell \rangle - \sum_\ell \langle I_\ell^{(s)} \rangle ,
\end{aligned} 
\end{equation}
with
\begin{equation}\label{Isl}
\begin{aligned}
\langle I_\ell^{(s)} \rangle &= - \sum_{ab} \left(\mathcal{L}_\ell\left[\rho^I(t) \right]\right)^{ab}  e^{-i (E_a - E_b)t}  \bigg( \sum_{j}\sum_s \langle j | c_s^\dagger c_s  \ket{\phi_a(t)} \bra{\phi_b(t)}  j \rangle \bigg).
\end{aligned}
\end{equation}

\section{Adding dissipation acting on the channel}

Additional Lindblad dissipation acting on the QD can be accounted for by adding to Eq.~(\ref{EQF}) a dissipator of the form
\begin{equation}
\begin{aligned}
\mathcal{D}_I[\rho(t)] &= \gamma_{I} \bigg(2 L \rho(t) L^\dagger
- L^\dagger L \rho(t)
-  \rho(t) L^\dagger L \bigg).
\end{aligned}
\end{equation}
In the Floquet basis and in interaction picture with respect to $H_\mathrm{QD}^\mathrm{eff}(t)$, this dissipator reads
\begin{equation}\label{Dincoh}
\begin{aligned}
(\mathcal{D}_{I}[\rho^I(t)])^{ab}
&= \gamma_{I}\sum_{cd}\sum_{kk'} \bigg[ 
2\left( e^{i (\Delta_{dbk'} + \Delta_{ack})t} L^{ack} L^{\dagger dbk'}\right) \rho^{I,cd}(t) \\
&\hspace{3cm}
- \left(  e^{i (\Delta_{cdk'} + \Delta_{ack})t} L^{\dagger ack} L^{cdk'} \right) \rho^{I,db}(t) 
- \left(  e^{i (\Delta_{dbk'} + \Delta_{cdk})t} L^{\dagger cdk} L^{dbk'}\right) \rho^{I,ac}(t) \bigg].
\end{aligned}
\end{equation}
By contrast with~(\ref{EQF}), there is no spectral dependence of the decay rates, as expected from Lindblad dissipation. By solving the master equation~(\ref{EQF}) after adding on its right-hand-side the dissipator~(\ref{Dincoh}), we can then calculate the associated particle current
\begin{equation}\label{IIncoh}
\langle I_{I}(t) \rangle =-\sum_{a,b} \left(\mathcal{D}_{I}\left[\rho^I(t) \right]\right)^{ab}  e^{-i (E_a - E_b)t} \bigg( \sum_{j}\sum_s \langle j | c_s^\dagger c_s  \ket{\phi_a(t)} \bra{\phi_b(t)}  j \rangle \bigg). \\
\end{equation}

\section{Numerical details}

Here we discuss some numerical details about the resolution of the master equation~(\ref{EQF}).

First, in order to write the master equation, we computed the quasienergies from Eq.~(\ref{EVeq}) using the procedure stated above with $N = 10000$, a number of time steps which insures precise enough values for the quasinergies. To avoid the divergence of the superconducting density of states~(\ref{DOS}), we modify their expression as
\begin{equation}\label{DOS2}
D_\ell(E) = \frac{|E|}{\sqrt{E^2 - \Delta_\ell^2}} \Theta( |E| - \Delta_\ell) \longrightarrow \frac{|E|}{\sqrt{E^2 - \Delta_\ell^2} + \epsilon^2} \Theta( |E| - \Delta_\ell)
\end{equation}
with a small parameter $\epsilon = 10^{-1}$. This procedure introduces finite values for the peaks of the DOS, and thus for the real part of the complex rates~(\ref{rates}). Their imaginary parts require another treatment to become finite. Here, we transform the denominators appearing in the integrations as 
\begin{equation}
\begin{aligned}
&\Omega_{\ell \pm}(E) \to  \frac{\gamma_\ell}{\pi} \,\mathrm{P.V.} \int_{-\Omega_f}^{\Omega_f} d\tilde{\omega}D_\ell(\tilde{\omega}) \mathrm{Re}\left[ \frac{n(\tilde{\omega})}{E + \tilde{\omega}+ i \epsilon} \right], \\
&\Omega_{\ell 1}(E) \to \frac{2\gamma_\ell}{\pi} \,\mathrm{P.V.} \int_{0}^{\Omega_f} d\tilde{\omega}D_\ell(\tilde{\omega}) u_{\ell}^{*}(\tilde{\omega})v_{\ell}^{*}(\tilde{\omega}) \mathrm{Re}\left( -\frac{n(\tilde{\omega})}{E + \tilde{\omega} + i \epsilon}  + \frac{1-n(\tilde{\omega})}{E - \tilde{\omega} + i \epsilon}  \right)
\end{aligned}
\end{equation}
and introduce a cutoff $\Omega_f = 100$ to the integration domain.

We solved the master equation using brute force resolution of the differential equations~(\ref{EQF}), with random or particular initial states, from a initial time $t_i = 0$ to a final time $t_f = 10 \gamma^{-1}$, where $\gamma \equiv \gamma_\ell = 10^{-2}$, in order to reach the steady state. We then averaged the particle currents over one period $T$. 

\section{QD energy dependence of the subgap currents}

In this section, we show how the positions of the MAR peaks depend on the QD energy $\omega$, which can be tuned experimentally by applying a gate voltage on the QD. Figure~\ref{figSM} shows the conductance $dI/dV$ as a function of both the bias voltage $V$ and the QD energy $\omega$, for vanishing or not Cooper-pair tunnelling amplitude $g \equiv g_\ell$. For $g = 0$, the conductance exhibit the standard diamond structure and vanish in the subgap region. For $g \neq 0$, subgap peaks appear and their positions depend on both the bias voltage and QD energy $\omega$. Hence, the subgap peaks do not just occur at rational fractions of the gap energy $\Delta$: they also depend on the properties of the QD.

\begin{figure}[tb]
\begin{center}
\includegraphics[width=0.7\textwidth]{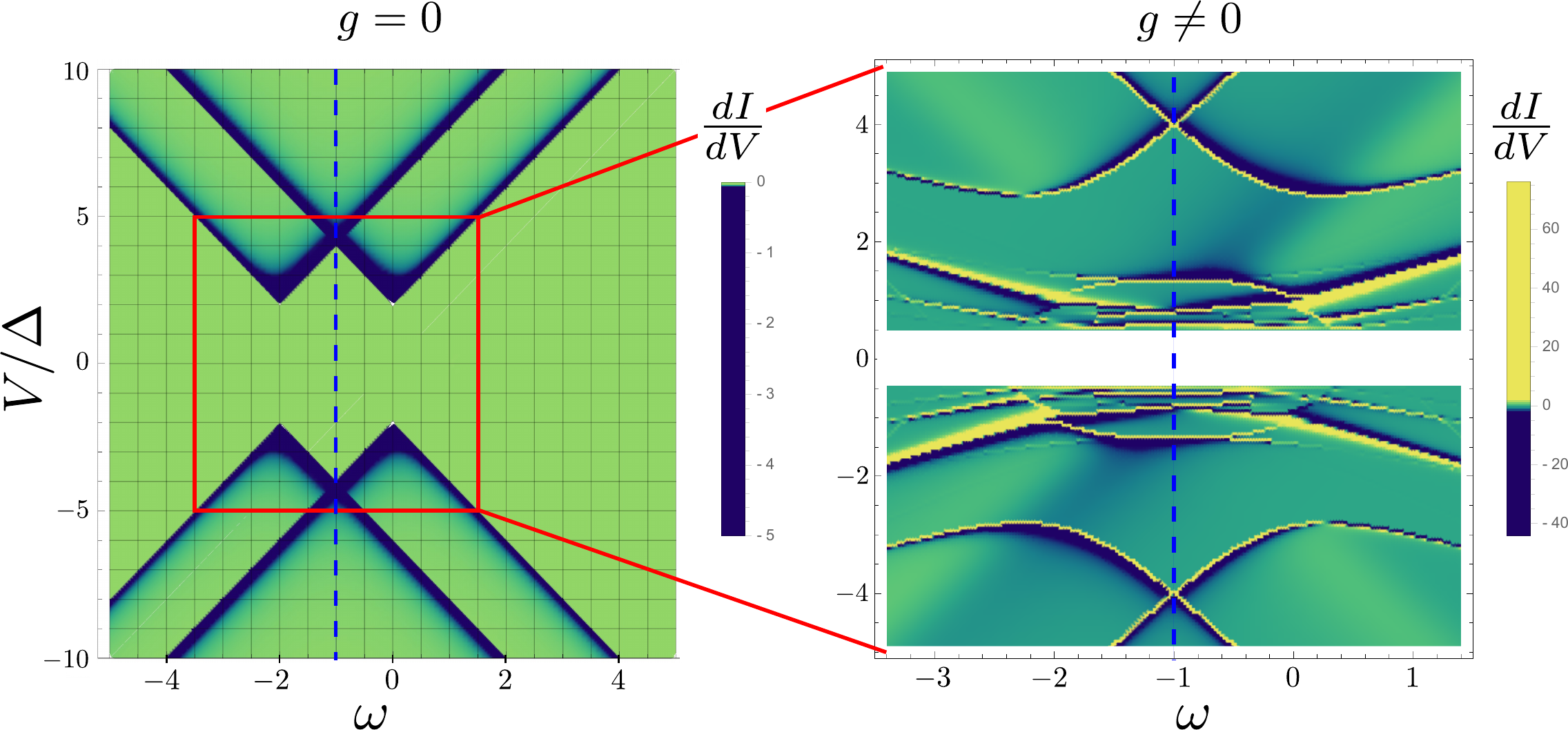}
\end{center}
\caption{Conductance $dI/dV$ corresponding to the particle current $I$ (in units of $\gamma$) in the right moderate-gap lead as a function of the bias voltage $V$ and QD energy $\omega$, for $g = g_\ell = 0$ (left) and $0.8$ (right). Other parameters are $U = 2$, $\gamma = 10^{-2}$ and $T_\ell = 0$, in units chosen so that $\Delta_\ell \equiv \Delta = 1$. MAR peaks appear for non-vanishing $g$, and their positions depend on both the source-drain voltage and QD energy $\omega$. The dashed blue line corresponds to the value of $\omega$ considered in the main text.}
\label{figSM}
\end{figure} 

\end{widetext}

\end{document}